\documentclass[conference]{IEEEtran}
\usepackage{cite}

\ifCLASSINFOpdf
   \usepackage[pdftex]{graphicx}
\else
\fi
\usepackage{array}
\usepackage{fixltx2e}
\usepackage{url}

\usepackage{caption} 
\usepackage{subcaption}
\usepackage{listings}
\usepackage{color}

\begin{document}
%
\title{A Parallel Task-based Approach to Linear Algebra}

\author{\IEEEauthorblockN{Ashkan Tousimojarad}
\IEEEauthorblockA{School of Computing Science\\
University of Glasgow\\
Email: a.tousimojarad.1@research.gla.ac.uk}
\and
\IEEEauthorblockN{Wim Vanderbauwhede}
\IEEEauthorblockA{School of Computing Science\\ University of Glasgow\\
Email: wim@dcs.gla.ac.uk}
}


%


\maketitle

\begin{abstract}
Processors with large numbers of cores are becoming commonplace.
In order to take advantage of the available resources in these systems, 
the programming paradigm has to move towards increased parallelism.
However, increasing the level of concurrency in the program does not
necessarily lead to better performance. 
Parallel programming models have to provide flexible ways of defining parallel tasks
and at the same time, efficiently managing the created tasks.
OpenMP is a widely accepted programming model for shared-memory architectures.
In this paper we highlight some of the drawbacks in the OpenMP tasking approach, and propose an alternative model based 
on the Glasgow Parallel Reduction Machine (GPRM) programming framework. 
As the main focus of this study, we deploy our model to solve a fundamental linear algebra problem, LU factorisation of
sparse matrices. We have used the SparseLU benchmark from the BOTS benchmark suite, and compared the results obtained
from our model to those of the OpenMP tasking approach. The TILEPro64 system has been used to run the experiments. The results are very promising, not only because of the performance improvement for this particular problem, but also because they verify the task management efficiency, stability, and flexibility of our model, which can be applied to solve problems in future many-core systems.
\end{abstract}


%
\IEEEpeerreviewmaketitle

\section{Introduction}

Task-based parallel programming models are evolving rapidly. With the emerge of many-core processors, they can compete with  data-parallel approaches, while offering more flexibility, because of their MIMD nature. A task is any form of computation that can be run in parallel with other tasks, if their data dependencies allow. Most of these programming models provide the programmer with some keywords to express parallelism in an imperative language such as C/C++. Pure functional programming languages on the other hand provide native parallelism, but compared to mainstream languages such as C++ and Java, none of them have found widespread adoption. Even if a many-core programming language would find wide adoption, it would in the short term obviously be impossible to rewrite the vast amount of single-core legacy code libraries, nor would it be productive. Our mission is therefore to propose a programming model that can be integrated into existing codes in imperative languages, while offering native parallelism, similar to functional languages. Before going into the details of our approach, the Glasgow Parallel Reduction Machine (GPRM), we would like to briefly review some of the available models in the market, namely Clojure, Chapel, Intel Cilk Plus, Intel Threading Building Blocks (TBB), and OpenMP.

Parallel programming is not as simple as sequential programming. In addition to \emph{what} to compute, the programmer should specify \emph{how} to coordinate the computation. Clojure \cite{hickey2008clojure} --also called a modern Lisp-- is a functional programming language that targets the Java Virtual Machine (JVM). Clojure's syntax is based on S-expressions, i.e. lists where the first element represents the operation and the other elements the operands.  GPRM uses a similar approach for the internal representation of its communication code. 

OpenCL \cite{stone2010opencl} is an industrial standard for heterogeneous architectures. It basically defines a set of core functionality that is supported by all devices, and allow vendors to expose more programming interfaces as well as hardware features. It is however not as easy to use as the following models for shared memory architectures.

Intel Cilk Plus which is based on the Cilk++\cite{leiserson2010cilk++} is an extension to C/C++ to provide both task and data parallelism. Is has become popular because of its simplicity. It has \texttt{\small \_Cilk\_spawn} and \texttt{\small \_Cilk\_sync} keywords to spawn and synchronise the tasks. \texttt{\small \_Cilk\_for} loop is a parallel replacement for sequential loops in C/C++. Intel Cilk Plus starts a pool of threads in the beginning of the program which is analogous to the GPRM thread pool.

Intel Threading Building Blocks (TBB) is another well-known approach for expressing task-based parallelism \cite{reinders2007intel}. Intel TBB is an object-oriented C++ runtime library that contains data structures and algorithms to be used in parallel programs. It abstracts the low-level threading details, which is similar to the GPRM design consideration. However, the tasking comes along with an overhead. Conversion of the legacy code to TBB requires restructuring certain parts of the program to fit the TBB templates. One of the advantages of TBB over OpenMP and Cilk Plus is that it does not require specific compiler support.

OpenMP can be called the most widely used programming standard in the shared-memory architectures. Since the release of OpenMP 3.0 \cite{ayguade2009design}, irregular parallelism can be expressed by means of the OpenMP \emph{task}s. Moreover, OpenMP works very well with predictable data parallel situations compared to Cilk Plus and TBB. This makes it a challenging competitor for new programming models such as GPRM. We have compared the performance of GPRM with that of OpenMP in 2 different scenarios: first a micro-benchmark which has structured parallelism, and second, a linear algebra problem which fits very well into less structured task-based parallelism.

It has been studied in \cite{duran2008adaptive, podobas2010comparison, teruel2009openmp, tousimojarad137glasgow} that OpenMP performs poorly for fine-grained tasks. It indicates that the programmer is responsible to figure out how a problem with specific input parameters would fit on a particular platform. As a common solution, programmers use a cutoff value when creating OpenMP tasks to avoid composing fine-grained tasks. Firstly, finding a proper cutoff value is not straightforward, and sometimes needs a comprehensive analysis of the program. It often depends on the application structure and the input data set \cite{duran2008evaluation}. Secondly, only in some special cases, such as recursion, the cutoff value can be controlled by the user code. Leaving the decision to the runtime system has been proposed as an alternative. The idea is to aggregate tasks by not creating some of the user specified tasks and instead executing them serially. Adaptive Task Cutoff (ATC) \cite{duran2008adaptive} implemented in the Nanos runtime system --a research OpenMP runtime system-- is a scheme to modify the cutoff dynamically based on profiling data collected early in the program's execution. This, however, cannot be done without any overhead at all, plus the scheme needs to be extended to find all interesting situations for cutoff. Instead, in GPRM there is a technique to specify which task would initially run on which thread. This reduces the overhead of tasks considerably. Moreover, there are some worksharing constructs that can be used along with tasks to avoid generating too many fine-grained tasks. 

The trade-off between task granularity and the number of tasks in OpenMP is covered in \cite{teruel2009openmp}. The authors suggest that tasks should have sufficient granularity and the granularity should be increased as the number of consumer threads increases. This is to ensure that all threads are kept busy doing some useful work. They have also explored that the number of tasks has an effect on the load balance, which means programmers have to trade-off between the number of tasks and granularity in order to get a fair load balance, hence a good performance.

For the purposes of this paper, we will show how OpenMP fails to operate normally for fine-grained tasks, while the proposed model copes with such situations naturally (See Section \ref{matrix-multiplication}). Furthermore, we will introduce a hybrid worksharing-tasking approach to avoid creating too many tasks (See Section Section \ref{sparselu-sec}). In other words, a hybrid methodology for exploiting both task and data parallelism will be employed to solve an LU Factorisation problem on a homogeneous many-core processor . 

Lower-Upper factorisation over sparse matrices is a fundamental linear algebra problem. Due to the sparseness of the matrix, conventional worksharing solutions are not enough, since a lot of load imbalance exists. As a well-known testcase, we have used the SparseLU benchmark from the the Barcelona OpenMP Tasks Suite (BOTS) \cite{duran2009barcelona}. In this problem, the matrix is organised in blocks that may not be allocated. More information along with the source code is openly available.

\section{GPRM }

The Glasgow Parallel Reduction Machine (GPRM) \cite{tousimojarad137glasgow} provides a task-based approach to many-core programming. The programmer structures programs into task code, written as C++ classes, and communication code, written in a restricted subset of C++. A \emph{task} is a list of bytecodes representing an S-expression, e.g. $(S_1 \,(S_2 \,10) \,20)$ represents a task $S_1$ taking two arguments, the first argument is the task $S_2$ which takes  as argument the numeric constant $10$,   and the second argument is the numeric constant 20. GPRM executes the corresponding list of bytecodes with concurrent evaluation of function arguments. For more details on the compilation of the S-expressions and the bytecode see \cite{tousimojarad137glasgow}.
In our parlance, a \emph{task node} consists of a \emph{task kernel} and a \emph{task manager}. A \emph{task kernel} is typically a complex, self-contained entity offering a specific functionality to the system, which on its own is not aware of the rest of the system. The \emph{task kernel} has run-to-completion semantics. The corresponding \emph{task manager} provides an interface to the kernel.  The computational \emph{task kernel}s are written as C++ classes. This means that the end user simply creates classes in the GPRM::Kernel namespace. 

Conceptually, GPRM consists of a set of \emph{tile}s connected over a network. Each \emph{tile} consists of a \emph{task node} and a FIFO queue for incoming packets. Every \emph{tile} runs in its own thread and blocks on the FIFO. The system is event driven, with two possible types of events: arrival of a packet, or the events generated by the \emph{task kernel}. The latter is either creation of a packet or modification of the local state. The reduction engine, i.e. the \emph{task manager} evaluates the bytecode via parallel dispatch of packets requesting computations to other tiles.

Threads in GPRM are treated as execution resources. Therefore, for each processing core there is a thread with its own \emph{task manager}. At the beginning, a pool of threads is created before the actual program starts. It has been pointed out in the Introduction that GPRM offers an effective way of combining tasks with worksharing constructs to avoid creating fine-grained tasks. For instance, instead of creating tasks in a loop which is common in the OpenMP tasking approach \cite{ayguade2009design}, one can create as many tasks as the concurrency level in GPRM, each of which with their own indices. These indices can be then used by a worksharing construct to specify which elements of the loop belongs to which thread. The concurrency level is defined as the number of jobs that can be theoretically run simultaneously in the system.

Normally, when the tasks are fairly equal, the best result can be obtained by choosing the concurrency level as the same as the number of threads, which is itself as the same as the number of cores in GPRM. Although in Section \ref{sparselu-sec} tasks are not equal, as a solution one can use the GPRM parallel loops to balance the load amongst threads. This solution, as will be shown, works very well when medium size or large sparse matrices are used.

\section{Parallel Loops}
We have created a number of useful parallel loop constructs for use in GPRM. These worksharing
constructs corresponds to the \texttt{\small for} worksharing construct in OpenMP, in
the sense that they are used to distribute different parts of a work
among different threads. However, there is a big difference in how they perform the operation. In OpenMP, the user marks a loop as an OpenMP \texttt{\small for} with a desirable scheduling strategy, and the OpenMP runtime decides which threads should run which part of the loop; in GPRM, multiple instances of the same task --normally as many as the concurrency level-- are generated, each with a different index (similar to the \texttt{\small global\_id} in OpenCL). Each of these tasks calls the parallel loop passing in their own index to specify which parts of the work should be performed by their host thread.

\begin{figure}[h]
\begin{centering}
\includegraphics[width=0.3\textwidth]{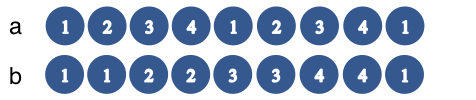}
\par\end{centering}
\caption{\label{worksharing} Partitioning a nested $m(3\times3)$ or a single $m(9)$ loop amongst $n(4)$ threads. a) Step size of 1, as in the \texttt{\small par\_for} and \texttt{\small par\_nested\_for}, b) Contiguous}
\end{figure}

The \texttt{\small par\_for} construct is used to parallelise a single loop. It distributes the work 
 in a Round-Robin fashion to the threads. A \texttt{\small par\_nested\_for} treats a nested loop as a single loop and follows the same pattern to distribute the work. Alternatively, the \emph{contiguous} method gives every thread an $m/n$ chunk, and the remainder $m\%n$ is distributed one-by-one to the foremost threads. These methods are shown in Fig \ref{worksharing}. The need to parallelise nested loops arises often, e.g. in situations where there are variable size loops such as  the SparseLU benchmark in Section \ref{sparselu-sec}.

\definecolor{mygreen}{rgb}{0,0.6,0}
\definecolor{mygray}{rgb}{0.5,0.5,0.5}
\definecolor{mymauve}{rgb}{0.58,0,0.82}

\lstset{ %
  aboveskip=15pt,
  belowskip=5pt,
  framextopmargin=2pt,
  framexbottommargin=2pt,
  linewidth=\columnwidth,
  xleftmargin=10pt,
  xrightmargin=5pt,
  backgroundcolor=\color{white},   
  basicstyle=\footnotesize,        
  breakatwhitespace=false,         
  breaklines=true,                 
  captionpos=b,                    
  commentstyle=\itshape\color{mygreen},    
  deletekeywords={...},            
  escapeinside={\%*}{*)},          
  extendedchars=true,              
  frame=lines,                    
  keepspaces=true,                 
  language=C++,                 
  morekeywords={*,...},            
  numbers=left,                    
  numbersep=2pt,                   
  numberstyle=\tiny, 
  rulecolor=\color{black},         
  showspaces=false,                
  showstringspaces=false,          
  showtabs=false,                  
  stepnumber=1,                    
  stringstyle=\color{mymauve},     
  tabsize=2,                       
  title=\lstname                   
}

\begin{lstlisting}[caption={Implementation of the \texttt{\small par\_for}}, label=parallel-for]
template<typename Tclass, typename Param1>
int par_for(int start,int size,int ind, int CL, Tclass* TC, int (Tclass::*work_function)(int,int,int,Param1), Param1 p1) {
/* ind: Index, CL: Concurrency Level */
 int turn=0;
 for(int i = start; i < size;) {
 	if(turn % CL == ind) {
	 (TC->*work_function(i,start,size,p1);
	 i = i + CL;
	}
	else {
	 i++;
	 turn++;
	}
 }
 return;
}
\end{lstlisting}

The \texttt{\small par\_for} and \texttt{\small par\_nested\_for} loops in GPRM are implemented using C++ templates and member-function pointers. The implementation of these worksharing constructs are given in Listing \ref{parallel-for} and \ref{parallel-nested-for}.  They would be our worksharing constructs by default.
The Contiguous parallel loops have similar implementations. We denote Contiguous parallel loops as Contiguous GPRM approaches.

Another useful worksharing construct is a parallel nested loop. Since the GPRM \texttt{\small par\_nested\_for}
is implemented with minimum overhead, it is a significantly useful
worksharing construct, as we will see in the next sections.

\begin{lstlisting}[caption={Implementation of the \texttt{\small par\_nested\_for}},label=parallel-nested-for]
template <typename Tclass, typename Param1>
int par_nested_for(int start1, int size1, int start2, int size2, int ind, int CL, Tclass* TC, int (Tclass::*work_function)(int,int,int,int,int,int,Param1), Param1 p1) {
 int turn=0;
 for(int i = start1; i < size1; i++) {
 	for(int j = start2; j < size2;) {
 	 if((turn >= 0) && (turn % CL == ind)) {
 	  (TC->*work_function)(i,j,start1,size1,start2,size2,p1);
	  j = j + CL;
	  if(j >= size2)	turn = size2 - j + ind; 
	 }
 	 else {
	  j++;
	  turn++;
	 }
 	}
 }
 return;
}
\end{lstlisting}

\section{Experimental Setup\label{experimental-setup}}

The Tilera TILEPro64 Tile Processor is a 32-bit VLIW multicore with 64 tiles, interconnected via multiple $8\times8$ mesh networks. It provides distributed cache-coherent shared memory by default. It has 16GB of DDR memory, 
but in order to use the global address space shared among all tiles, addressing is limited to 32-bit, i.e. 4GB. It has per-core L1 caches
of 8KB, and L2 caches of 64KB. The union of all L2 caches across the chip comprises the distributed L3 cache. The operating frequency of the cores is 866MHz. Out of 64 tiles, one is used for the PCI communication, and the other 63 tiles have been used for our experiments. For our experiments in this study, we have used the \emph{tile-g++} compiler provided by MDE 3.0 from the Tilera Corporation, and is based on the GCC version 4.4.3. The compiler flags -O2 and -std=c++0x have been specified. It is worth stating that the TILEPro64 runs \emph{Tile Linux} which is based on the standard open-source Linux version 2.6.36.

\section{Matrix Multiplication Micro-benchmark\label{matrix-multiplication}}

In this section, we use a naive matrix multiplication algorithm with a triple nested loop as a micro-benchmark 
to evaluate the overhead of the OpenMP approach compared to the model we use to solve matrix problems. The code is given at Listing \ref{mmcode}. 

As our aim is to use this micro-benchmark to identify the most important barriers on the way, we change the interpretation of the problem to performing multiple jobs. Suppose that the product of an $m \times n$ matrix A and an $n \times p$ matrix B is the $m \times p$ matrix C. We want to parallelise the first loop of the triple nested loop, which loops on $m$, therefore $m$ becomes the number of jobs for this problem. The size of each job is identified by the sizes of the next two loops in the triple nested loop, i.e.  $p*n$. We have chosen $n=p$ to make the problem more regular. We end up with matrices with the following specification: $A: m \times n$, $B: n \times n$, and $C: m \times n$.  Due to the poor data locality of this algorithm, one should not expect to see a linear speedup.

\begin{lstlisting}[caption={Matrix Multiplication Micro-benchmark}, label=mmcode]
for (int i = 0; i < m; i++){
 for(int j = 0; j< p; j++){
  for(int k = 0; k <n;  k++){
   C[i*p+j] += A[i*n+k] * B[k*p+j];
  }
 }
}
\end{lstlisting}

Four approaches are selected for the comparison: I) The OpenMP \texttt{\small for} worksharing construct, II) The OpenMP \texttt{\small for} with \emph{dynamic} schedule and \emph{chunk\_size} of 1, III) The OpenMP Tasks, and IV) The GPRM \texttt{\small par\_for} construct.

\begin{figure*}
\begin{centering}
	\begin{subfigure}{0.65\columnwidth}
	\includegraphics[width=\columnwidth]{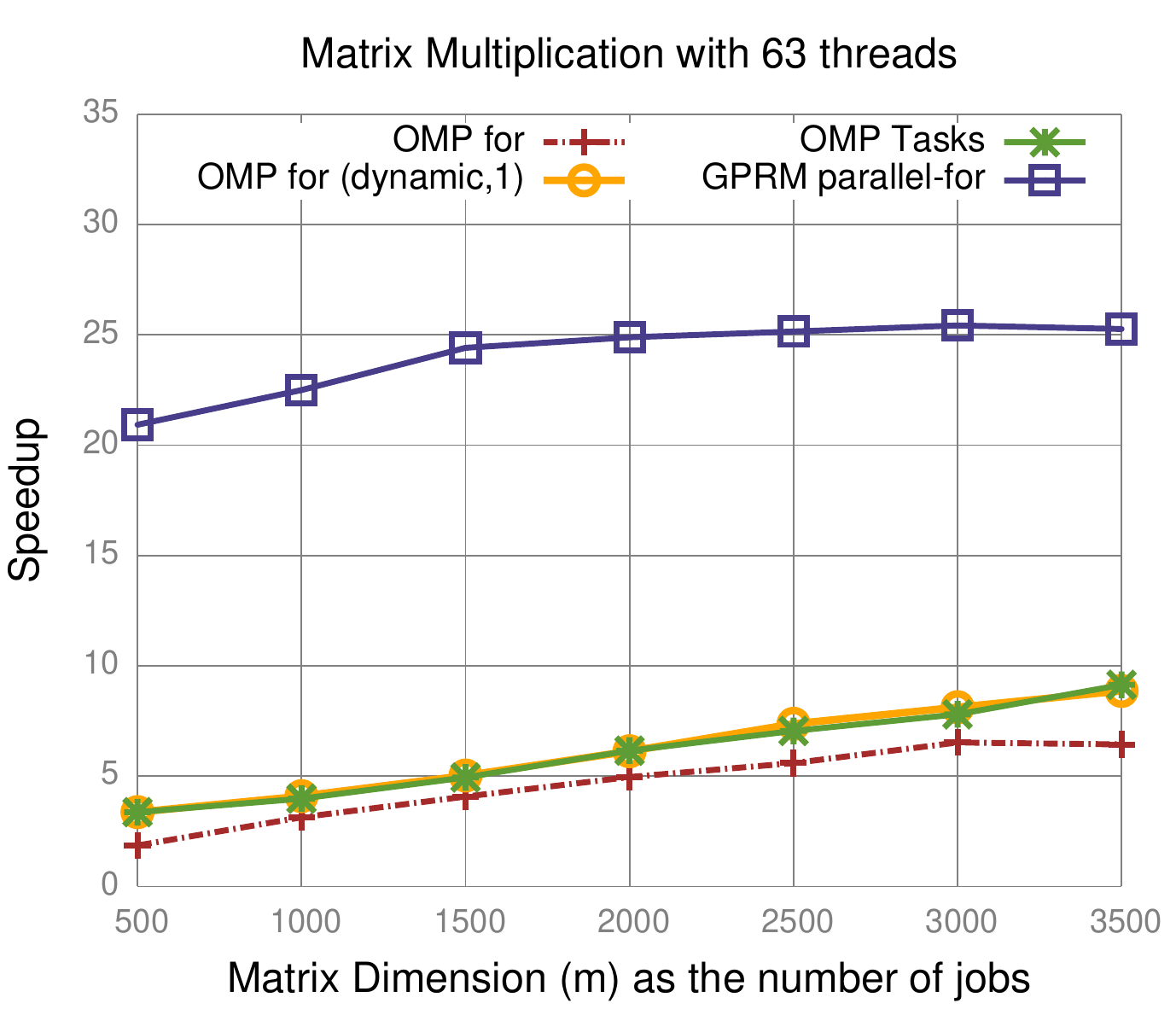}
	\caption{\label{matmul500} $500 \times 500$ - Small jobs}
	\end{subfigure}
	\hfill
	\begin{subfigure}{0.65\columnwidth}
	\includegraphics[width=\columnwidth]{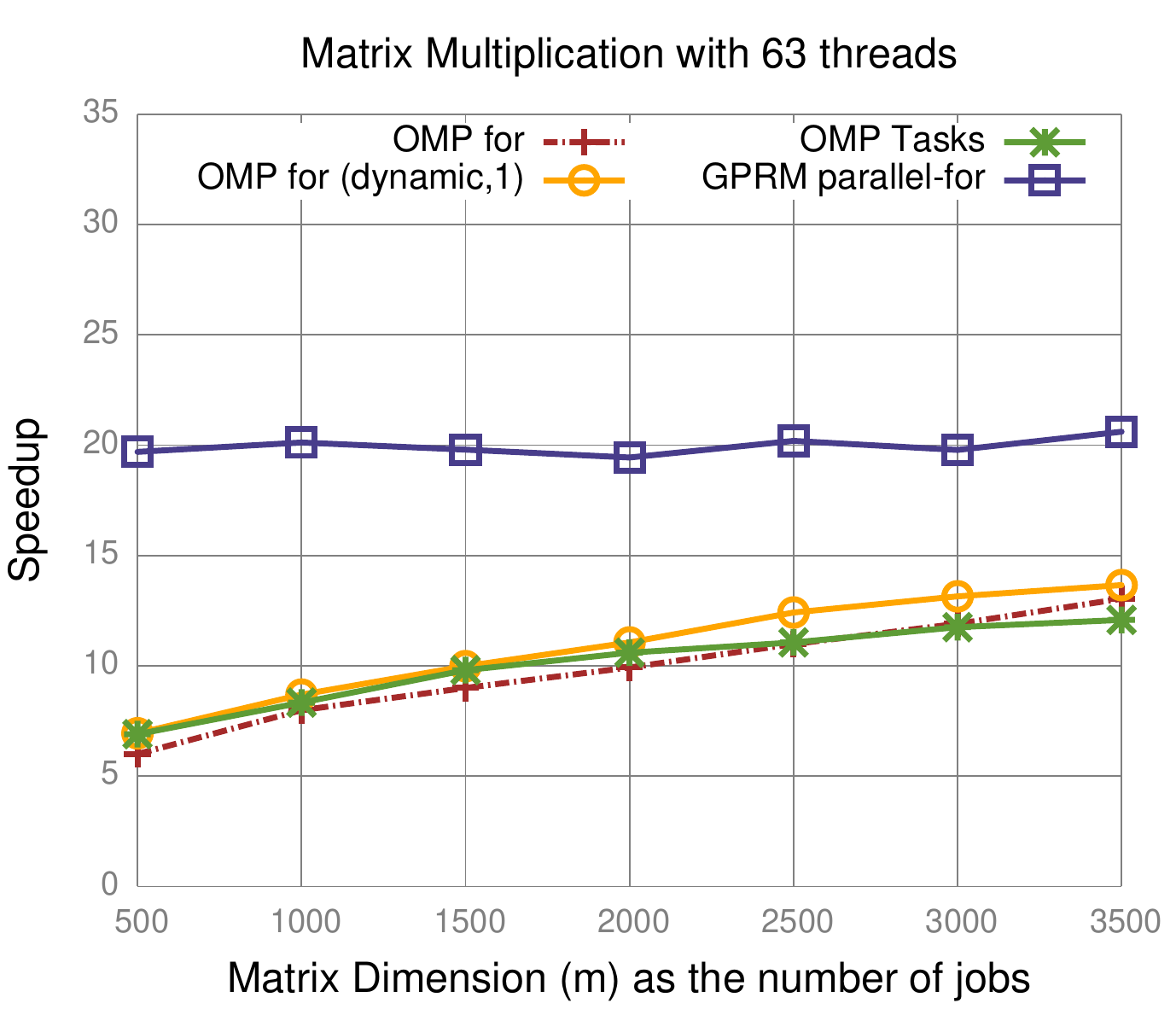}
	\caption{\label{matmul1000} $1000 \times 1000$ - Medium jobs}
	\end{subfigure}
	\hfill
	\begin{subfigure}{0.65\columnwidth}
	\includegraphics[width=\columnwidth]{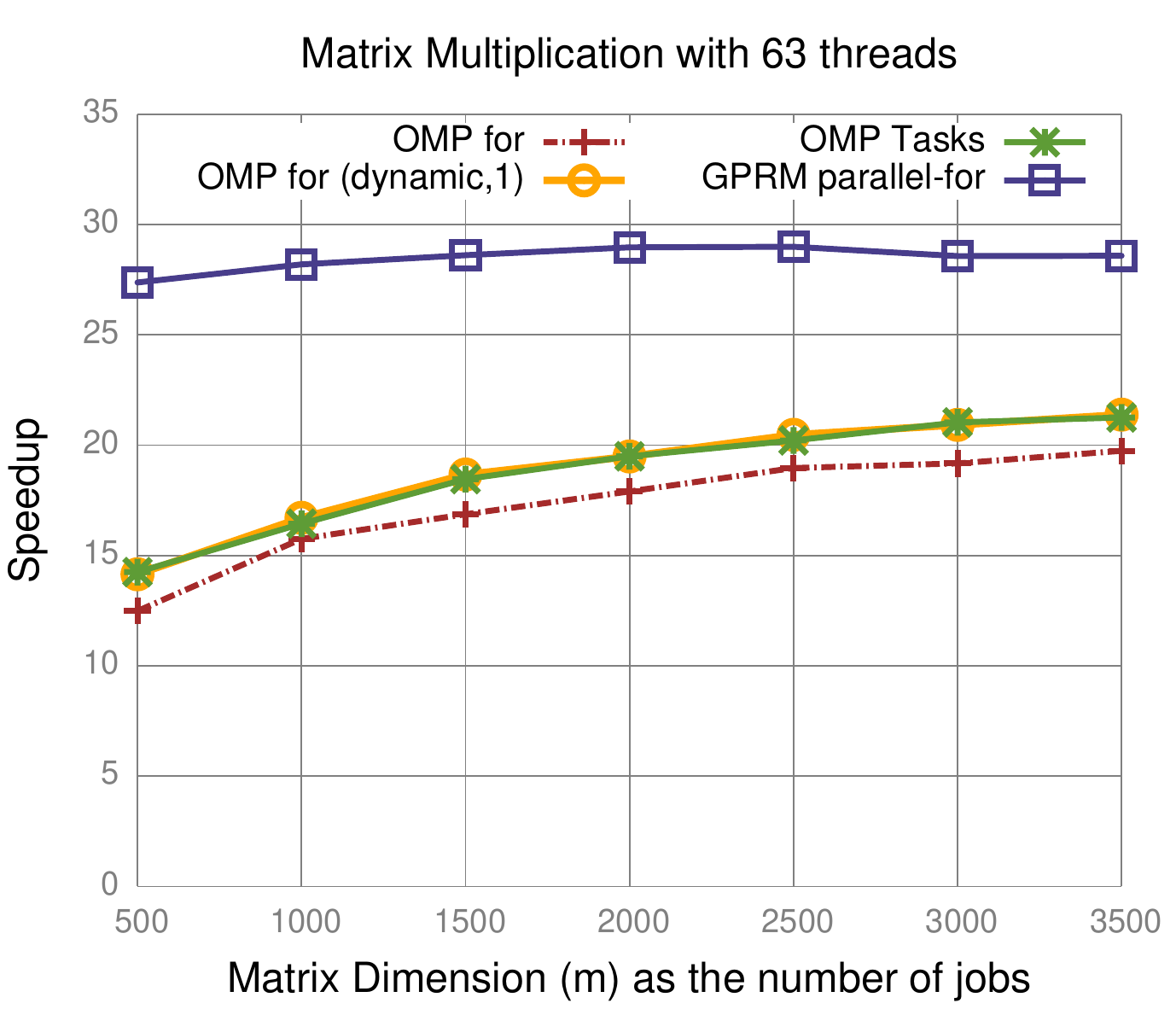}
	\caption{\label{matmul1500} $1500 \times 1500$ - Large jobs}
	\end{subfigure}
\caption{Matrix Multiplication: GPRM \texttt{\small par\_for} vs. different OpenMP approaches}
\label{matmulfig}
\end{centering}
\end{figure*}

Figure \ref{matmulfig} shows the performance measurement for different job sizes. GPRM outperforms OpenMP in all cases but especially for the small job case (even then, the job size is still not small enough to show the real overhead of having fine-grained tasks in OpenMP). To our best understanding, the performance difference is due to the overhead of thread scheduling,  which is more visible in the small job cases with short execution times.

In order to investigate the effect of task granularity on the behaviour of the OpenMP\rq{}s performance, we decrease the size of the tasks even more. We also looked at the influence of the cutoff value on the performance. Since the behaviours of the OpenMP worksharing constructs were fairly similar, only the default \texttt{\small omp for}  is used for the next experiment.

\begin{figure}[h]
\begin{centering}
\includegraphics[width=0.45\textwidth]{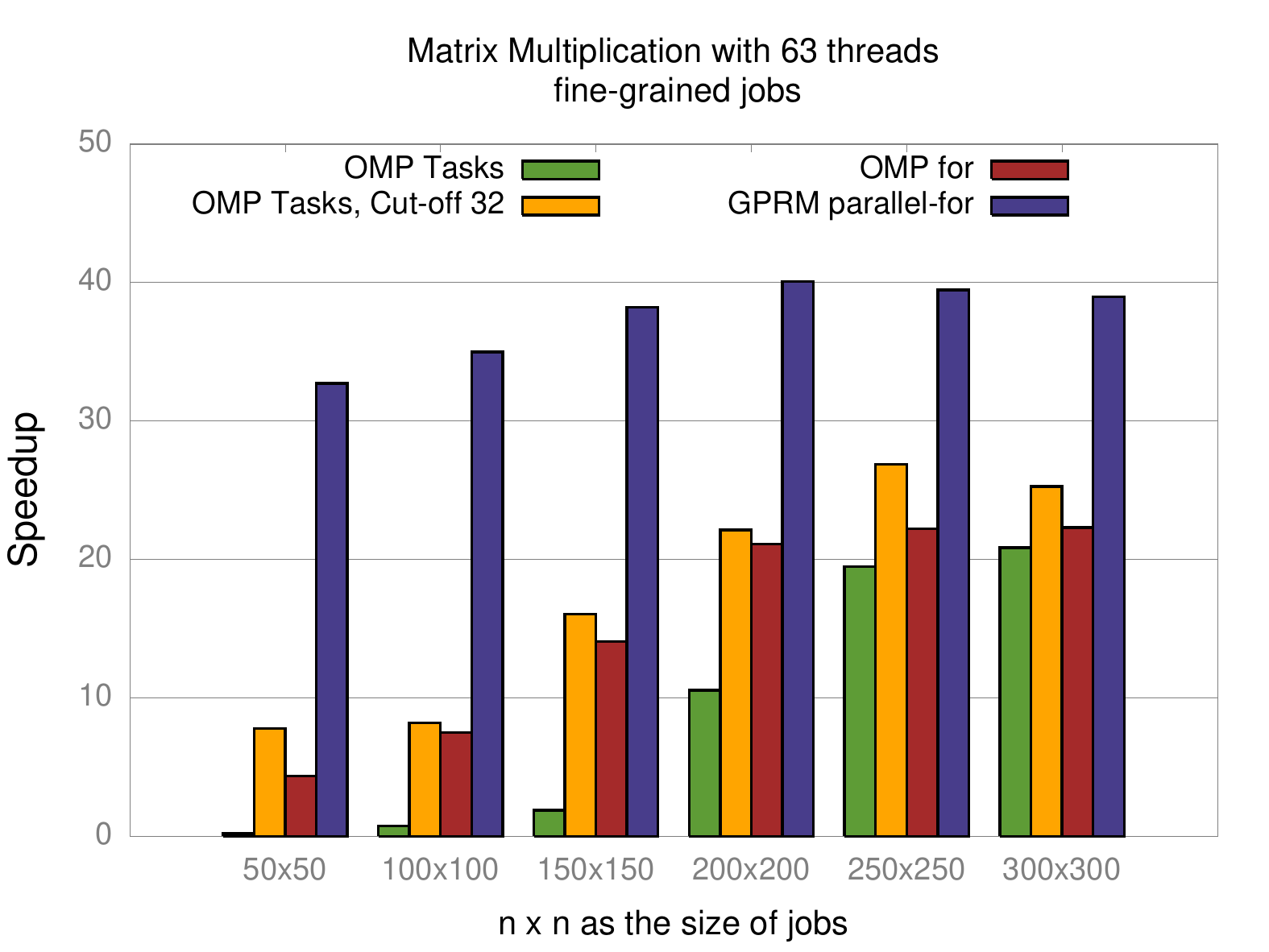}
\par\end{centering}
\caption{\label{matmulcutoff} Speedup measurement for fine-grained jobs, Number of the jobs: 200,000}
\end{figure}

To improve the behaviour of the tasking approach, we added a cutoff value for the tasks, such that only \emph{m/cutoff} tasks were created. This is similar to sequencing multiple tasks. Fig \ref{matmulcutoff} compares the speedup of a tuned version of the OpenMP task-based model with the other alternatives. We believe that the regularity of GPRM in assigning tasks to its threads and the lower overhead of its tasks make it the winner.

\begin{lstlisting}[caption={Matrix Multiplication Micro-benchmark with a cutoff value}, label=mmcutoff]
for (int i = 0; i < (m/cutoff); i++){
#pragma omp task firstprivate(i)
 for(int t = 0; t < cutoff; t++) {
  for(int j = 0; j < p; j++){ // p=n
   for(int k = 0; k < n; k++){ 
    C[(i*cutoff+t)*p+j] += A[(i*cutoff+t)*n+k] * B[k*p+j];
   }
  }
 }
}
\end{lstlisting}

Figure \ref{matmulcutoff} shows that the poor behaviour of fine-grained tasks can be remedied to a considerable extent by using a proper cutoff value. The OpenMP approaches gradually becomes better when the size of the job is increased. We have chosen the first two cases to show the effect of using a cutoff in more detail, because these cases show degraded performance compared to the sequential implementation if no cutoff is used at all. 

\begin{figure}[h]
\begin{centering}
\includegraphics[width=0.42\textwidth]{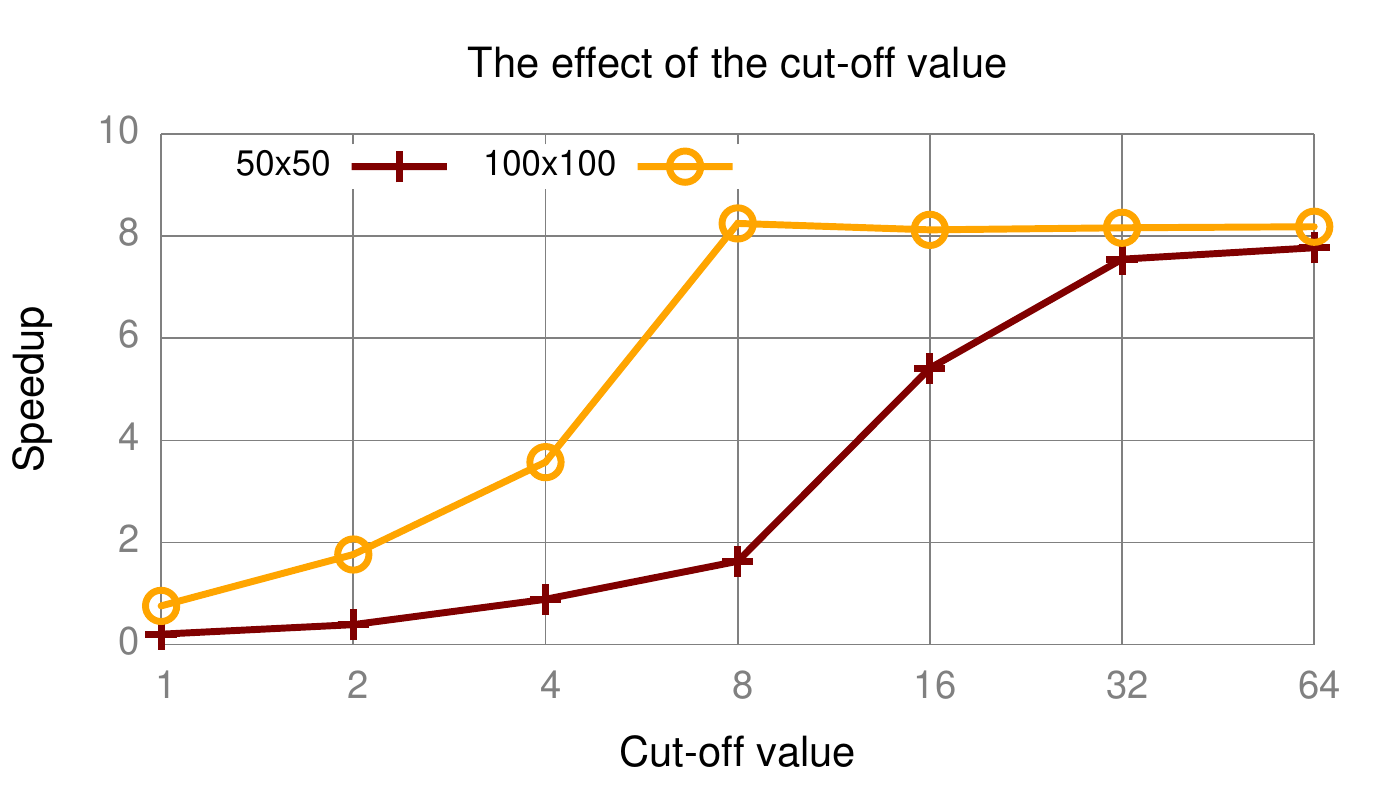}
\par\end{centering}
\caption{\label{taskcutoff} Speedup improvement by using a cutoff value for the fine-grained OpenMP tasks.
Number of the jobs: 200,000. Size of the jobs: 50$\times$50 - 100$\times$100}
\end{figure}

Figure \ref{taskcutoff} shows that a good choice of the cutoff value gives the speedup of 38.6$\times$ against the case with no cutoff and 7.8$\times$ against the sequential version, for the job size of $50\times50$ with 63 threads. The speedup for the job size of $100\times100$ is also improved by 10.8$\times$ compared to the case with no cutoff and 8.2$\times$ compared to the sequential runtime.

\section{Sparse LU Factorisation\label{sparselu-sec}}

\begin{figure}[h]
\begin{centering}
\includegraphics[width=0.50\textwidth]{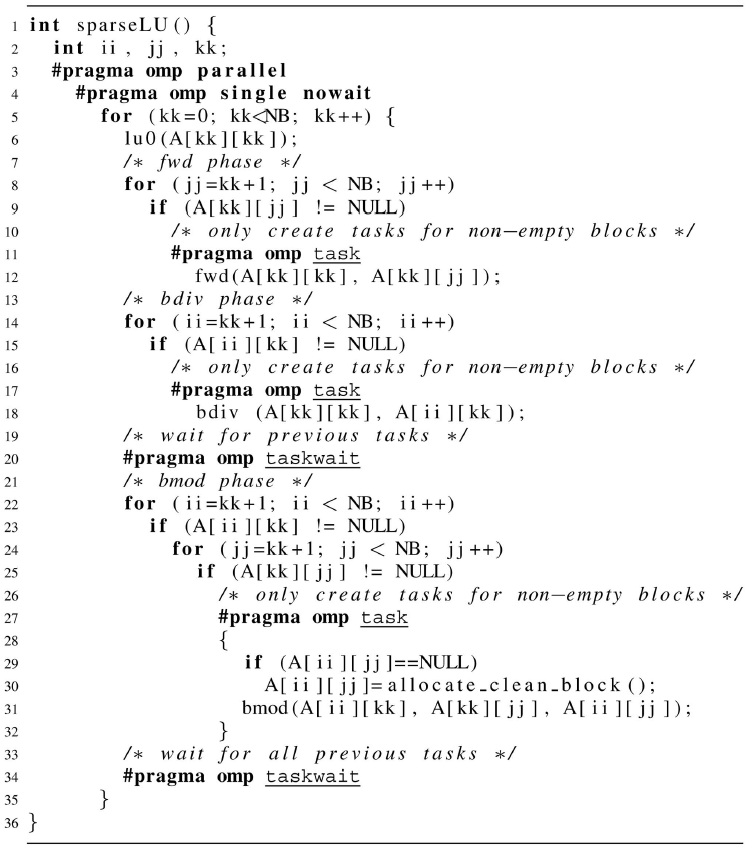}
\par\end{centering}
\caption{\label{sparselucopied} Main code of the SprseLU benchmark, without revealing OpenMP programming details \cite{ayguade2009design}}
\end{figure}

The SparseLU benchmark from the BOTS suite, which computes an LU factorisation over sparse matrices, is a proper example of matrix operations with load imbalance. In the OpenMP approach, a task is created for each non-empty block. The main SparseLU code from \cite{ayguade2009design} (omitting the details of the OpenMP task-based programming, such as dealing with shared and private variables) is copied in Fig \ref{sparselucopied}. The number of tasks and the granularity of them depends on the number of non-empty blocks and the size of each block, hence a cutoff value cannot be defined inside the user-written OpenMP code. As has also been discussed in \cite{ayguade2009design}, using OpenMP tasks results in better performance compared to using the \texttt{\small for} worksharing construct with dynamic scheduling. Therefore, we use the tasking approach for the comparison.

The code for the SparseLU benchmark in GPRM can be found in Listing \ref{SparseLU GPRM}.  

\begin{lstlisting}[caption={SparseLU code in GPRM, Concurrency Level: 63},label=SparseLU GPRM]
#include <GPRM/API.h>
using namespace GPRM::API;
GPRM::Kernel::SpLU sp;

void fwd_bdiv_tasks (int kk,float** A,
		 const int CL) {
#pragma gprm unroll
 for (int n = 1; n < (CL/2); n++) { 
  sp.fwd_t(kk, A, n-1, CL/2);  // fwd task
  sp.bdiv_t(kk, A, n-1, CL/2); // bdiv task
 }
}

void bmod_tasks (int kk,float** A,const int CL) {
#pragma gprm unroll
 for(int n = 1; n < CL; n++) {  
  sp.bmod_t(kk, A, n-1, CL);	// bmod task
 }
}

void GPRM::Compute_LU () { 
#pragma gprm seq
 { /* GPRM evaluates in parallel unless otherwise stated */
  float** A = init_task(); 
  for(int kk=0,kk<NB,i++)  { // NB: #Blocks
#pragma gprm seq  
   { 
    lu0_task(kk,A);
    fwd_bdiv_tasks(kk,A,63)
    bmod_tasks(kk,A,63)
   }
  }
 }
 return; }
\end{lstlisting}

The \texttt{\small unroll} pragma results in compile-time evaluation of any control construct in which any of the variable in the argument list occurs. In the example, this means that the \texttt{\small for} loop will be unrolled.
By default, expressions in non-kernel GPRM code are evaluated in parallel. The \texttt{\small seq} pragma forces sequential evaluation of the block it precedes.

It is worth mentioning that we have not changed the initialisation phase of generating sparse matrices in the BOTS benchmark suite. The matrices become sparser as the number of blocks increases. For example, in the case of 50$\times$50 blocks, the matrices are $85\%$ sparse, while for the cases with 100$\times$100 blocks, the matrices become $89\%$ sparse.

To obtain a fair distribution of the matrix elements amongst different threads  --bearing in mind the sparseness of the matrix--  we have used a \texttt{\small par\_nested\_for}, because the numbers of iterations are not fixed in this problem. 
The loops become smaller as \emph{kk} grows, which means that using a \texttt{\small par\_for} would, after a few iterations when 
\emph{outer\_iters$\,>\,$concurrency\_level}, lead to starvation of some of the threads. By using a 
\texttt{\small par\_nested\_for} the threads can get some work as long as the \emph{outer\_iters$*$inner\_iters$ \,>\, $concurrency\_level}. Therefore, in order to implement the \emph{fwd}, \emph{bdiv}, and \emph{bmod} tasks, one can use the GPRM APIs for the parallel loops, as shown in Listing \ref{phases}.

\begin{lstlisting}[caption={Implementation of the member functions of the SparseLU class. Work functions can also be defined similar to the phases in \cite{ayguade2009design}},label=phases]
#include <GPRM/API.h>
using namespace GPRM::API;

int SpLU::fwd_t(int kk,float** A,int ind,int CL) 
{return par_for(kk+1,NB,ind,CL,this,
		&SpLU::fwd_work,A);}

int SpLU::bdiv_t(int kk,float** A,int ind,int CL) 
{return par_for(kk+1,NB,ind,CL,this,
		&SpLU::bdiv_work,A);}

int SpLU::bmod_t(int kk,float** A,int ind,int CL) 
{return par_nested_for(kk+1,NB,kk+1,NB,ind,CL,
		this,&SpLU::bmod_work,A);}

/* The fwd function here is the same as fwd in the cited paper */
int SpLU::fwd_work (int jj,int kk,int NB,
		float** A) { 
 if(A[(kk-1)*NB+jj] != NULL)
  fwd(A[(kk-1)*NB+kk-1],A[(kk-1)*NB+jj]);
 return 0;
}

\end{lstlisting}

\begin{figure}[h]
\begin{centering}
\includegraphics[width=0.42\textwidth]{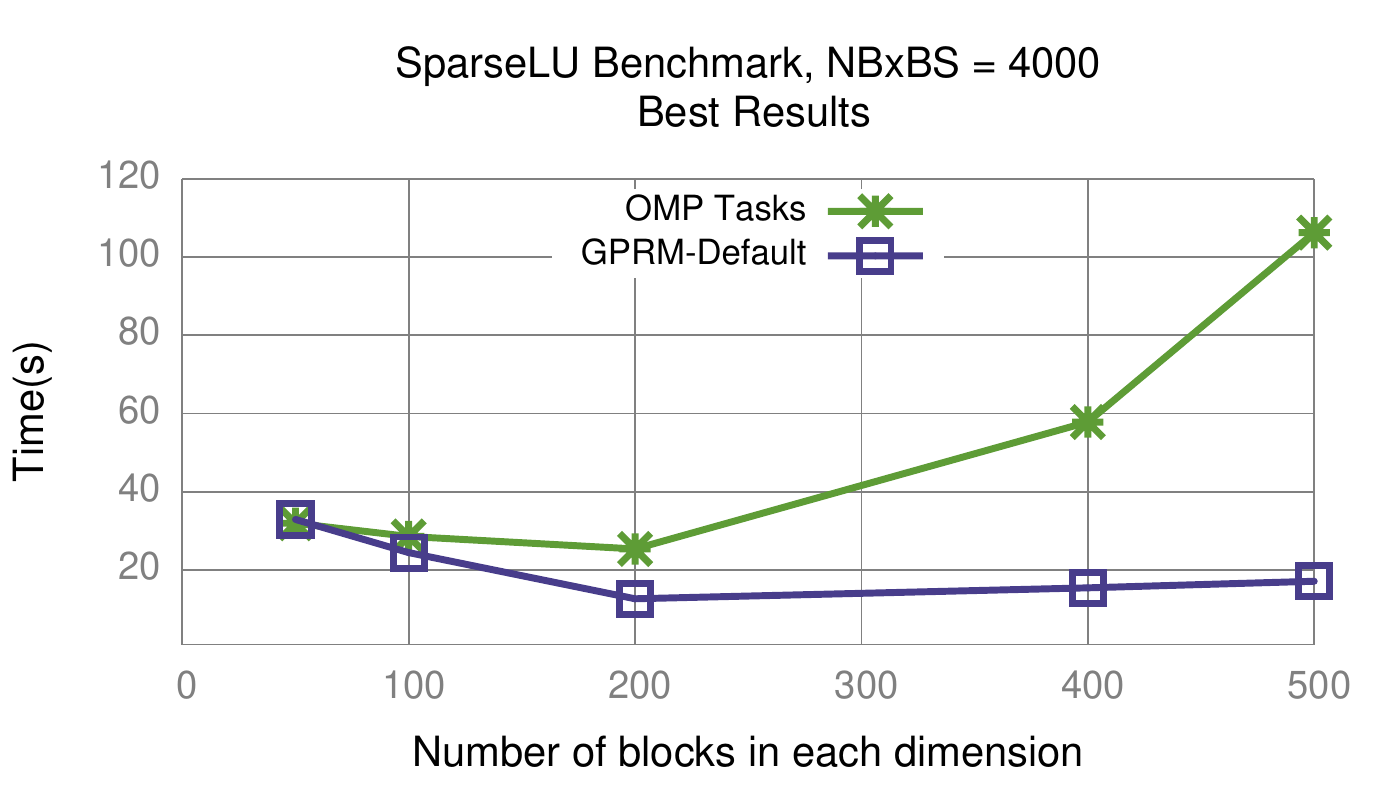}
\par\end{centering}
\caption{\label{sparselu4000} Execution time of sparse matrices of size 4000 with variable block sizes}
\end{figure}

Figure \ref{sparselu4000} shows a sparse matrix of 4000$\times$4000 divided into blocks of varying size. It is again clear that with larger numbers of blocks in each dimension which results in smaller block sizes, the OpenMP\rq{}s performance drops drastically. GPRM can deal with tiny 8$\times$8 blocks 6.2$\times$ better than the best result obtained by OpenMP.

Table \ref{bestresults} reveals that the best results for the OpenMP approach is not obtained with the default setting, which is as many threads as the number of cores. 
Besides the large difference in execution times when the block size becomes less than 20$\times$20, there is a significant performance degradation if the number of threads is set to the default value of 63. For example, the execution time becomes 12.25$\times$ worse than the best time for the last case. However, it is clear that GPRM reaches its best execution time without the need to tune the number of threads --here, the number of threads and concurrency level are the same for GPRM--. This is again due to the fact that instead of creating very small tasks, GPRM offers an efficient way of distributing the work amongst threads. 

\begin{figure*}
\begin{centering}
	\begin{subfigure}{1\columnwidth}
	\includegraphics[width=\columnwidth]{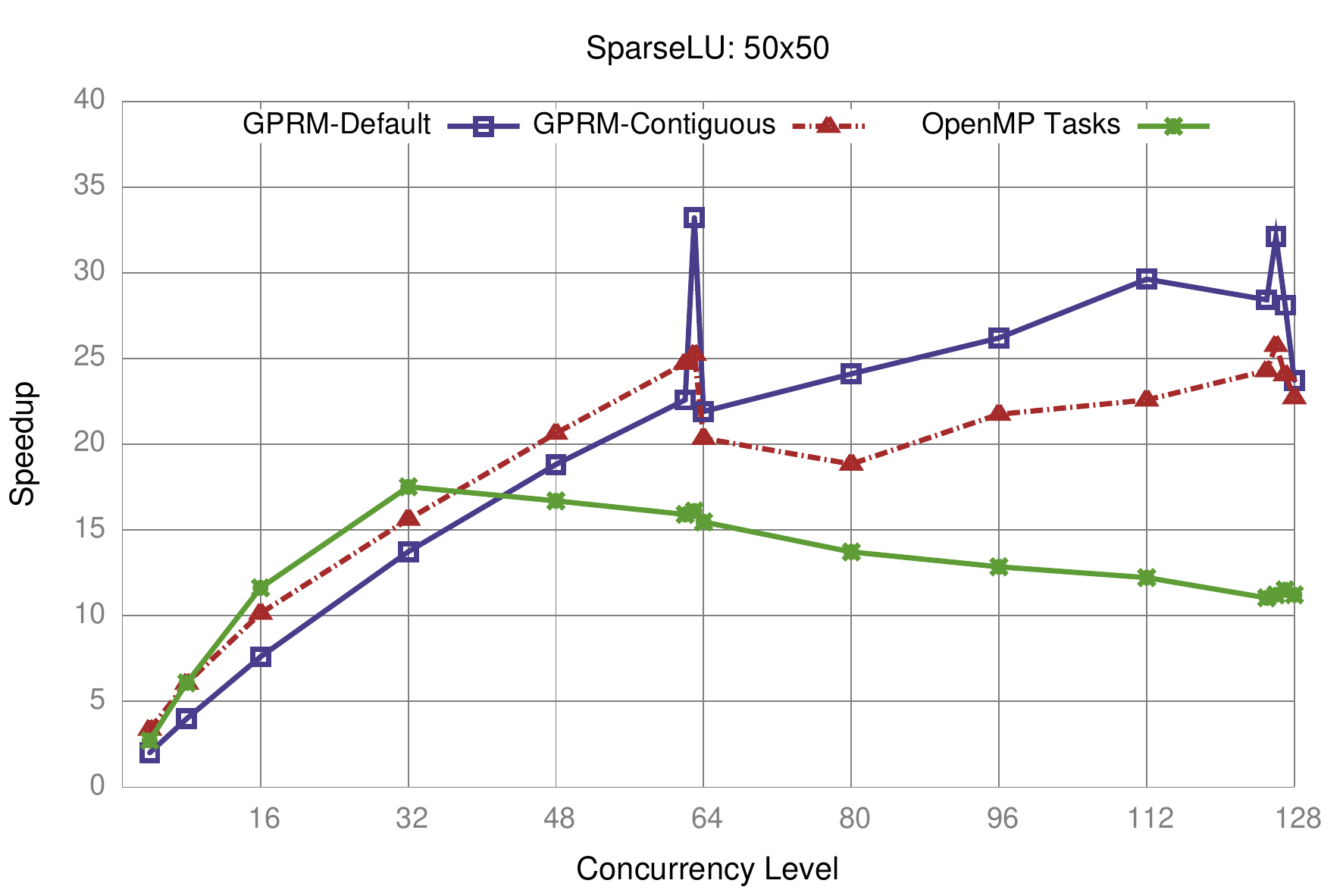}
	\caption{\label{sparselu50x50} $50 \times 50$}
	\end{subfigure}
	\hfill
	\begin{subfigure}{1\columnwidth}
	\includegraphics[width=\columnwidth]{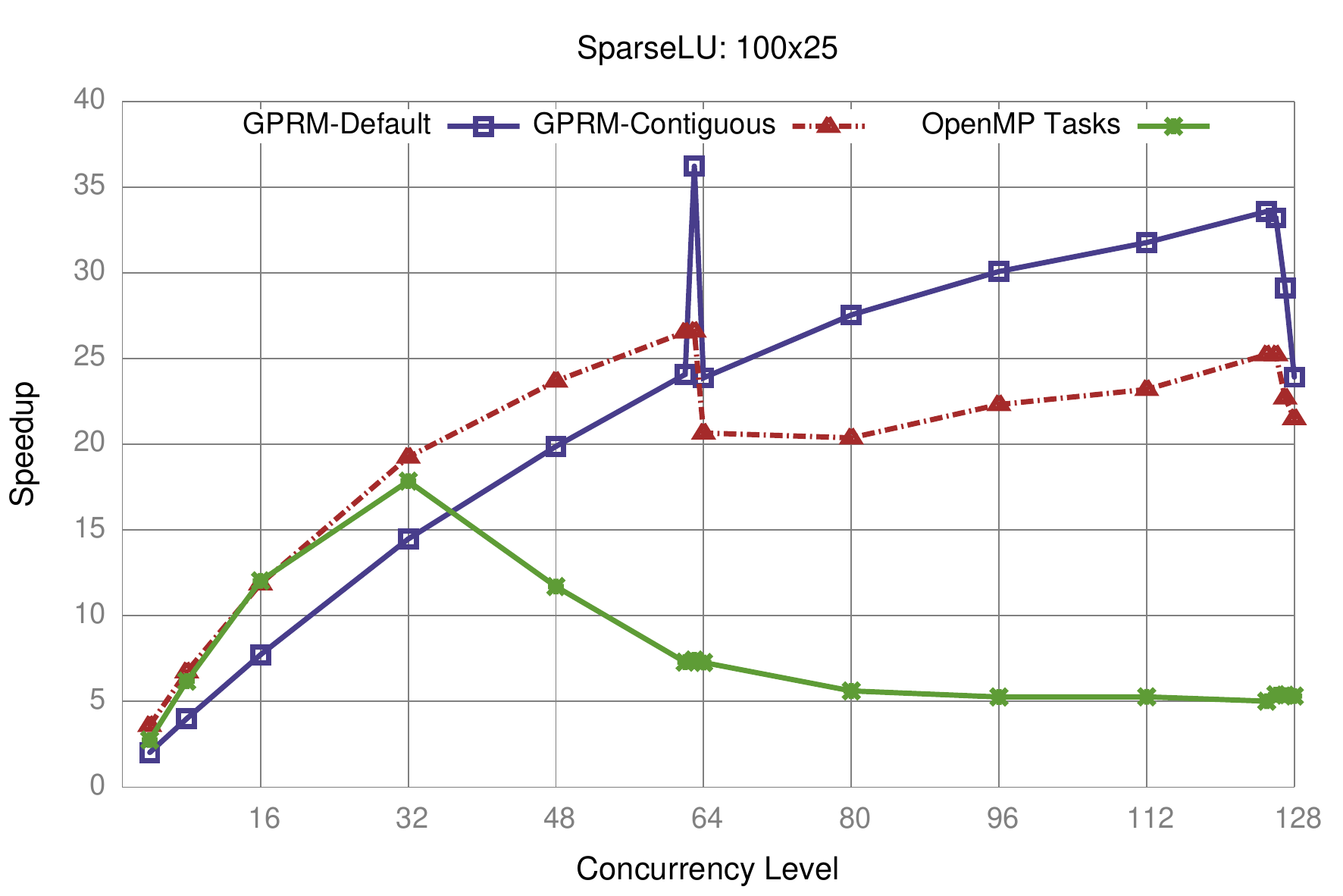}
	\caption{\label{sparselu100x25} $100 \times 25$}
	\end{subfigure}
\caption{SparseLU Factorisation: GPRM approaches vs. OpenMP Tasks, on the TILEPro64 with 63 available cores}
\label{sparselufig}
\end{centering}
\end{figure*}

The speedup diagrams for the SparseLU benchmark have been shown in Fig \ref{sparselufig}. We increased the concurrency level up to 128 to show how regular our approach is, in the sense that it gets its best performance with the factors of the number of cores. This is not surprising, since the problem has been partitioned amongst threads regularly, and therefore they can exploit the underlying architecture more efficiently. Since the OpenMP tasking model is different from ours, we simply increased the number of threads in that case.

\begin{table}[ht]
\renewcommand{\arraystretch}{1.3}
\caption{Number of threads for the best results}
{\centering
\resizebox{1\columnwidth}{!}{
\tabcolsep=0.20cm
\scalebox{0.4} {
\begin{tabular}{|l l l l l l|}
\hline 
Number of Blocks  &50&100&200&400&500\tabularnewline
\hline 

OpenMP tasks &64& 63& 32 &16 & 8\tabularnewline

GPRM &63 & 63 & 63 & 63 & 63\tabularnewline
\hline 
\end{tabular}
\label{bestresults}
}
}
}
\end{table}

\section{Discussions}

\subsection{OpenMP Performance Bottlenecks}

We have identified a number of performance bottlenecks when programming with OpenMP. The first is the thread migration overhead. This overhead can often be removed by statically mapping (pinning) the OpenMP threads to the execution cores. Using static thread mapping (pinning) in a platform with per-core caches could be very useful, particularly for load balanced data parallel problems, in which the portion of the work to be done by each thread is fairly equal, and therefore CPU time and local caches can be effectively utilised. Our study \cite{tousimojarad2013efficient} shows that for such a platform, static thread mapping is often a good practice in single-program environments, but for multiprogramming environments, in which different programs compete for the resources, it is not always efficient. We refer the readers to \cite{tousimojarad2013efficient, mazouz2011performance, tousimojarad2013cache} for detailed discussions on thread mapping (pinning) for OpenMP programs.

Another barrier is to find a proper cutoff point to avoid creating overly fine-grained tasks. Programmers have to be very careful about the granularity of the tasks, otherwise the results  maybe totally unexpected. It has also been observed that there is no guarantee that running an OpenMP program with the maximum number of threads --equal to the number of cores-- will result in the best performance.

\subsection{Comparison of OpenMP and GPRM Approaches}
For the SparseLU problem, although creating OpenMP tasks for non-empty blocks is a smart solution, it is not working very well for all matrices. The first
reason is that a single thread explores the whole matrix and creates
relatively small tasks for non-empty blocks, while in the proposed
solution implemented in GPRM, multiple threads look into their
portions of the work in parallel. The difference in performance can be noticeable especially
in the \emph{bmod} phase with a nested loop. As also reported in \cite{olivier2011scheduling}, combining the OpenMP \texttt{\small for} worksharing 
construct with tasks, as implemented in \texttt{\small sparselu\_for} in the BOTS benchmark suite, is not a viable approach with  OpenMP 3.0.

Secondly, in GPRM, every thread
has specific work based on the program\rq{}s \emph{task description} file. If needed, runtime
decisions will be applied to improve the performance, while OpenMP
creates the tasks dynamically and all decisions are taken dynamically
at runtime, which makes its task management less efficient when the
numbers of tasks and/or threads become larger. Moreover, the overhead of task management becomes significant when the tasks become more fine-grained, as made clear by the Matrix Multiplication micro-benchmark.

In GPRM, using a \texttt{\small par\_nested\_for}
construct in order to partition the matrix
in a non-contiguous manner results in a good load balance. Furthermore, our hybrid worksharing-tasking  technique is pretty much straightforward. There is no pressure on the programmer to worry about private and shared variables, in contrast to many other parallel programming models, including OpenMP.

The proposed programming toolkit, GPRM is stable, scalable, low-overhead, and flexible compared to OpenMP. Stable, as there is no need to change the default configurations, such as the number of threads in order to get the optimal performance. It scales as expected, which is showing continuous speedup as the concurrency level increases up to the number of cores. The overhead of task management is negligible, even for fine-grained tasks. It is flexible, because the number of tasks can be easily controlled by the programmer. It is also straightforward to specify which task to be run on which thread initially. If required, the runtime system can change the host thread dynamically.

\section{Conclusion}

Many-core systems have emerged to solve the existing problems faster. Giving the same amount of work to more execution resources simply means having more fine-grained tasks. In this paper, we have proposed a regular task composition approach which can address both regular and irregular parallel problems. We have shown that it fits naturally to the systems with per-core caches, which consequently makes it very promising for future many-cores. 

In this paper, our new model is used to solve a fundamental linear algebra problem, namely lower-upper factorisation over sparse matrices. We used a matrix multiplication micro-benchmark to highlight the differences between the proposed model and OpenMP. For the small jobs, GPRM outperformed OpenMP by 2.8$\times$ to 11$\times$. 
For the medium-sized jobs, the speedup  improvement ranged  from 1.5$\times$ to 3.3$\times$. As the jobs get larger the difference became less significant. For the large jobs, the speedup ranged from 1.3$\times$ to 2.2$\times$. 

As a real-world example we used the SparseLU testcase from the BOTS benchmark suite. We used a sparse matrix of 4000$\times$4000 divided into blocks of varying size. We demonstrated that for the larger numbers of blocks, i.e. smaller block sizes, the difference is considerable. The main advantage of  GPRM is that it does not need to be tuned it terms of the number of threads. By contrast, tuning is crucial for OpenMP, otherwise for fine-grained tasks  a huge drop in performance is inevitable. 

We also investigated the impact of the concurrency level on speedup. Again, GPRM scaled 2$\times$ better than the best result obtained using OpenMP for both the 50$\times$50 and 100$\times$100 cases. For \emph{concurrency level=63}, which is the default setting, the speedup improvements were respectively 2.1$\times$ and 4.9$\times$.

As GPRM can offer a hybrid task-data parallelism, utilising it on hybrid CPU-GPU systems remains an open research area.


\section*{Acknowledgment}

The first author would like to thank the Scottish Informatics and Computer Science Alliance (SICSA) for supporting his research studies.



%


\bibliographystyle{IEEEtran}
\bibliography{ISPDC2014}

\begin{thebibliography}{10}
\providecommand{\url}[1]{#1}
\csname url@samestyle\endcsname
\providecommand{\newblock}{\relax}
\providecommand{\bibinfo}[2]{#2}
\providecommand{\BIBentrySTDinterwordspacing}{\spaceskip=0pt\relax}
\providecommand{\BIBentryALTinterwordstretchfactor}{4}
\providecommand{\BIBentryALTinterwordspacing}{\spaceskip=\fontdimen2\font plus
\BIBentryALTinterwordstretchfactor\fontdimen3\font minus
  \fontdimen4\font\relax}
\providecommand{\BIBforeignlanguage}[2]{{%
\expandafter\ifx\csname l@#1\endcsname\relax
\typeout{** WARNING: IEEEtran.bst: No hyphenation pattern has been}%
\typeout{** loaded for the language `#1'. Using the pattern for}%
\typeout{** the default language instead.}%
\else
\language=\csname l@#1\endcsname
\fi
#2}}
\providecommand{\BIBdecl}{\relax}
\BIBdecl

\bibitem{hickey2008clojure}
R.~Hickey, ``The clojure programming language,'' in \emph{Proceedings of the
  2008 symposium on Dynamic languages}.\hskip 1em plus 0.5em minus 0.4em\relax
  ACM, 2008, p.~1.

\bibitem{stone2010opencl}
J.~E. Stone, D.~Gohara, and G.~Shi, ``Opencl: A parallel programming standard
  for heterogeneous computing systems,'' \emph{Computing in science \&
  engineering}, vol.~12, no.~3, p.~66, 2010.

\bibitem{leiserson2010cilk++}
C.~E. Leiserson, ``The cilk++ concurrency platform,'' \emph{The Journal of
  Supercomputing}, vol.~51, no.~3, pp. 244--257, 2010.

\bibitem{reinders2007intel}
J.~Reinders, \emph{Intel threading building blocks: outfitting C++ for
  multi-core processor parallelism}.\hskip 1em plus 0.5em minus 0.4em\relax
  O'Reilly Media, Inc., 2007.

\bibitem{ayguade2009design}
E.~Ayguad{\'e}, N.~Copty, A.~Duran, J.~Hoeflinger, Y.~Lin, F.~Massaioli,
  X.~Teruel, P.~Unnikrishnan, and G.~Zhang, ``The design of openmp tasks,''
  \emph{Parallel and Distributed Systems, IEEE Transactions on}, vol.~20,
  no.~3, pp. 404--418, 2009.

\bibitem{duran2008adaptive}
A.~Duran, J.~Corbal{\'a}n, and E.~Ayguad{\'e}, ``An adaptive cut-off for task
  parallelism,'' in \emph{High Performance Computing, Networking, Storage and
  Analysis, 2008. SC 2008. International Conference for}.\hskip 1em plus 0.5em
  minus 0.4em\relax IEEE, 2008, pp. 1--11.

\bibitem{podobas2010comparison}
A.~Podobas and M.~Brorsson, ``A comparison of some recent task-based parallel
  programming models,'' in \emph{Proceedings of the 3rd Workshop on
  Programmability Issues for Multi-Core Computers,(MULTIPROG'2010), Jan 2010,
  Pisa}, 2010.

\bibitem{teruel2009openmp}
X.~Teruel, C.~Barton, A.~Duran, X.~Martorell, E.~Ayguad{\'e}, P.~Unnikrishnan,
  G.~Zhang, and R.~Silvera, ``Openmp tasking analysis for programmers,'' in
  \emph{Proceedings of the 2009 Conference of the Center for Advanced Studies
  on Collaborative Research}.\hskip 1em plus 0.5em minus 0.4em\relax IBM Corp.,
  2009, pp. 32--42.

\bibitem{tousimojarad137glasgow}
A.~Tousimojarad and W.~Vanderbauwhede, ``The glasgow parallel reduction
  machine: Programming shared-memory many-core systems using parallel task
  composition,'' \emph{EPTCS}, vol. 137, pp. 79--94.

\bibitem{duran2008evaluation}
A.~Duran, J.~Corbal{\'a}n, and E.~Ayguad{\'e}, ``Evaluation of openmp task
  scheduling strategies,'' in \emph{OpenMP in a new era of parallelism}.\hskip
  1em plus 0.5em minus 0.4em\relax Springer, 2008, pp. 100--110.

\bibitem{duran2009barcelona}
A.~Duran, X.~Teruel, R.~Ferrer, X.~Martorell, and E.~Ayguade, ``Barcelona
  openmp tasks suite: A set of benchmarks targeting the exploitation of task
  parallelism in openmp,'' in \emph{Parallel Processing, 2009. ICPP'09.
  International Conference on}.\hskip 1em plus 0.5em minus 0.4em\relax IEEE,
  2009, pp. 124--131.

\bibitem{tousimojarad2013efficient}
A.~Tousimojarad and W.~Vanderbauwhede, ``An efficient thread mapping strategy
  for multiprogramming on manycore processors,'' in \emph{International
  Conference on Parallel Computing (ParCo 2013)}.\hskip 1em plus 0.5em minus
  0.4em\relax IOS Press, 2013, pp. 63--71.

\bibitem{mazouz2011performance}
A.~Mazouz, S.~Touati, and D.~Barthou, ``Performance evaluation and analysis of
  thread pinning strategies on multi-core platforms: Case study of spec omp
  applications on intel architectures,'' in \emph{High Performance Computing
  and Simulation (HPCS), 2011 International Conference on}.\hskip 1em plus
  0.5em minus 0.4em\relax IEEE, 2011, pp. 273--279.

\bibitem{tousimojarad2013cache}
A.~Tousimojarad and W.~Vanderbauwhede, ``Cache-aware parallel programming for
  manycore processors,'' \emph{ACM SIGARCH Computer Architecture News},
  vol.~41, 2013.

\bibitem{olivier2011scheduling}
S.~L. Olivier, A.~K. Porterfield, K.~B. Wheeler, and J.~F. Prins, ``Scheduling
  task parallelism on multi-socket multicore systems,'' in \emph{Proceedings of
  the 1st International Workshop on Runtime and Operating Systems for
  Supercomputers}.\hskip 1em plus 0.5em minus 0.4em\relax ACM, 2011, pp.
  49--56.

\end{thebibliography}

\end{document}